\documentclass[twocolumn, showpacs, nofootinbib]{revtex4-1}

\usepackage{amsmath}
\usepackage{amsfonts}

\begin{document}

\title{Conserved quantities in isotropic loop quantum cosmology}

\author{D. Cartin}
\email{cartin@naps.edu}
\affiliation{Naval Academy Preparatory School, 197 Elliot Street, Newport, Rhode Island 02841}

\begin{abstract}

We develop an action principle for those models arising from isotropic loop quantum cosmology, and show that there is a natural conserved quantity $Q$ for the discrete difference equation arising from the Hamiltonian constraint. This quantity $Q$ relates the semi-classical limit of the wavefunction at large values of the spatial volume, but opposite triad orientations. Moreover, there is a similar quantity for generic difference equations of one parameter arising from a self-adjoint operator.

\end{abstract}

\pacs{04.60.Pp, 98.80.Qc, 02.30.Xx}

\date{\today}

\maketitle

\section{Introduction}

Loop quantum gravity (LQG) has been a successful research program, in the sense that it is a rigorously derived picture of the behavior of the gravitational field in the quantum regime. However, it is difficult to apply as is to physical problems -- such as how quantum gravity effects in the realm of classical singularities affect cosmological models in general -- because of its complexity. This has led to the development of loop quantum cosmology (LQC), a symmetry-reduced version of LQG. The Hamiltonian constraint equations of LQC are fundamentally discrete, due to the corresponding nature of quantum geometry coming from LQG, and thus are {\it difference} equations, rather than differential equations. The study of such difference equations has allowed the use of various analytic solution techniques, e.g. generating functions~\cite{CarKhaBoj04}, but these quickly become difficult to work with, leaving only numerical methods~\cite{BriCarKha11}. Thus, it is fruitful to look for other analytic methods to obtain exact information about solutions to the quantum constraint equation; we do so by considering the discrete analogue of classical mechanics.

Discrete Lagrangian mechanics has been developed independently several times, in many contexts; for a list of previous work, see Marsden and West~\cite{MarWes01}. Specifically for LQC, these ideas were developed by Shojai and Shojai~\cite{ShoSho06}, but oriented more towards finding approximate solutions to difference equations. There is also the program of consistent discretizations~\cite{BahGamPul11}, aiming to solve discretized versions of constrained mechanics, so that the constraints are preserved under evolution of the system in the model parameter $n$ (for an isotropic model). There, the emphasis is on the algebraic consistency of the discretized mechanics. In this work, we discuss a discrete version of the familiar Lagrangian mechanics, and show that for the isotropic models currently used in LQC, there is a conserved quantity $Q$ -- in the sense that this function $Q_n$ of the single (discrete) parameter $n$ arising in the quantum Hamiltonian constraint equation is constant, regardless of what $n$ is evaluated at.

This paper is organized as follows. We start with an action principle for a functional ${\cal L}_n$ of a one-parameter sequence $s_n$, and derive the discrete version of the Euler-Lagrange equation which extremizes the action. Then, assuming the action is invariant under an infinitesimal transformation of $s_n$, we find a quantity $Q$ that remains constant under the action of the difference equation for $s_n$. As we will show, the generic Lagrangian for the isotropic models will be phase invariant under $s_n \to s_n' = s_n \exp(i \alpha)$; this generates a discrete version of the familiar conserved quantity for a phase invariant field $\phi(t)$. After this groundwork is laid, we turn to specific examples in isotropic LQC, and find the corresponding conserved quantities $Q$. We also consider general self-adjoint difference operators (or difference equations of the same form) and show they similarly have a conserved quantity.

\section{An action principle for difference equations}

For the rest of this work, we define the difference operator $\Delta$ acting on a sequence $s_n$ as
\begin{equation}
	\Delta s_n = s_n - s_{n-1}.
\end{equation}
One particularly useful identity is the discrete version of integration by parts, namely the equivalence
\begin{equation}
\label{IBP}
	\sum_{n = -M + 1} ^M F_n \Delta G_n = \sum_{n = -M + 1} ^M [\Delta (F_{n + 1} G_n) - G_n \Delta F_{n + 1}].
\end{equation}
Here the summation acts as the discrete analogue of a integral for functions of continuous variables. Much like the integral of a total derivative becomes a boundary term, the sum of a total difference gives a similar result:
\begin{equation}
\label{bound-term}
\sum_{n = -M + 1} ^M \Delta (F_{n + 1} G_n) = F_{M+1} G_M - F_{-M+1} G_{-M}.
\end{equation}
The ``unbalanced" limits in these series come from the definition of $\Delta$. These limits have also been chosen to be symmetric around $n = 0$, but this is not a requirement; all of the results obtained in this work would be similar for a different choice of limits.

Now we briefly derive the Euler-Lagrange equation for a one-parameter sequence, both to keep the discussion self-contained, and to develop the notation used here. We start with the action
$$
	S = \sum_{n = -M+1} ^M {\cal L}_n = \sum_{n = -M+1} ^M {\cal L} (n; s_n, \Delta s_n).
$$
The format of ${\cal L}_n$ is to match that of ``standard" continuous actions, i.e. with a coordinate variable $q$ and its time derivative ${\dot q}$. We will assume that the boundary values $s_M$ and $s_{-M}$ are fixed data, and determine the rest of the sequence. First, we seek the general Euler-Lagrange equation for the sequence $s_n$ that extremizes the value of $S$. In this vein, we consider an infinitesimal variation of the sequence $s_n$, namely,
$$
	s_n \to s_n' = s_n + \alpha \eta_n,
$$
for $|n| < M$, and find the sequence such that
$$
	\frac{dS}{d \alpha} \biggl |_{\alpha = 0} = \sum_{n = -M+1} ^M \frac{d {\cal L} (n; s_n', \Delta s_n')}{d \alpha} \biggl |_{\alpha = 0} = 0.
$$
This gives
\begin{equation}
\label{ext-1}
	\frac{d {\cal L}_n}{d \alpha} = \frac{\partial {\cal L}_n}{\partial s_n} \eta_n + \frac{\partial {\cal L}_n}{\partial (\Delta s_n)} (\Delta \eta_n).
\end{equation}
When placed inside the summation, the last term in the equation (\ref{ext-1}) can be integrated by parts as in (\ref{IBP}), i.e.
\begin{eqnarray}
\nonumber
	&& \sum_{n = -M+1} ^M \frac{\partial {\cal L} (n; s_n, \Delta s_n)}{\partial (\Delta s_n)} (\Delta \eta_n) \\
\nonumber
	&& = \sum_{n = -M+1} ^M \Delta \biggl[ \frac{\partial {\cal L} (n+1; s_{n + 1}, \Delta s_{n + 1})}{\partial (\Delta s_{n + 1})} \eta_n \biggl] \\
\label{parts-eqn}
	&& - \sum_{n = -M+1} ^M \Delta \biggl[ \frac{\partial {\cal L} (n+1; s_{n + 1}, \Delta s_{n + 1})}{\partial (\Delta s_{n + 1})} \biggr] \eta_n.
\end{eqnarray}
From equation (\ref{bound-term}), the total difference in (\ref{parts-eqn}) becomes
\begin{eqnarray}
	\sum_{n = -M+1} ^M && \hspace{-0.25in} \Delta \biggl[ \frac{\partial {\cal L} (n+1; s_{n + 1}, \Delta s_{n + 1})}{\partial (\Delta s_{n + 1})} \eta_n \biggl]  \\
	\nonumber
	&=& \frac{\partial {\cal L} (n; s_n, \Delta s_n)}{\partial (\Delta s_n)} \biggl|_{n = M + 1} \eta_M \\
	\nonumber
	&-& \frac{\partial {\cal L} (n; s_n, \Delta s_n)}{\partial (\Delta s_n)} \biggl|_{n = - M + 1} \eta_{-M}
	\nonumber
	= 0.
\end{eqnarray}
where the last equality results from the assumption that the boundary values are fixed, so $\eta_M = \eta_{-M} = 0$. Thus,
\begin{eqnarray*}
	\frac{dS}{d \alpha} \biggl |_{\alpha = 0} &=& \sum_{n = -M+1} ^{M - 1} \biggl[ \frac{\partial {\cal L} (n; s_n, \Delta s_n)}{\partial s_n}	\\
	&-& \Delta \biggl( \frac{\partial {\cal L} (n+1; s_{n + 1}, \Delta s_{n + 1})}{\partial (\Delta s_{n + 1})}\biggl) \biggl] \eta_n.
\end{eqnarray*}
The remaining values $\eta_n$ are arbitrary, so this gives us the Euler-Lagrange equations\footnote{We note here that another form of this equation has been derived in the literature. If the original discrete Lagrangian is of the form ${\cal L} (n; s_n, s_{n - 1})$ -- that is, all appearances of $s_{n - 1}$ in ${\cal L}_n$ are not necessarily part of a difference $\Delta s_n$, then the Euler-Lagrange equations are of the form
$$
	\frac{\partial {\cal L} (n; s_n, s_{n - 1})}{\partial s_n} + \frac{\partial {\cal L} (n + 1; s_{n + 1}, s_n)}{\partial s_n} = 0,
$$
for $|n| \le M - 1$. Note that the position of the variables in ${\cal L}_n$ is important in the above equation, since ${\cal L}_n (u, v) \ne {\cal L}_n (v, u)$. This is the form usually used in the literature, such as by Bahr, Gambini and Pullin~\cite{BahGamPul11}, and Shojai and Shojai~\cite{ShoSho06}, although the latter allows the Lagrangian to be a function of $s_{n -2}$ as well as $s_n, s_{n - 1}$. If ${\cal L}$ is a function of $s_n$ and $\Delta s_n$ only, however, this is equivalent to the equation (\ref{EL-eqn}).}
\begin{equation}
\label{EL-eqn}
	\frac{\partial {\cal L} (n; s_n, \Delta s_n)}{\partial s_n} -  \Delta \biggl[ \frac{\partial {\cal L} (n+1; s_{n + 1}, \Delta s_{n + 1})}{\partial (\Delta s_{n + 1})} \biggl] = 0,
\end{equation}
for $|n| \le M - 1$.

As an example that will be relevant later, suppose we consider the action
\begin{equation}
\label{iso-action}
	S = \sum_{n = -M + 1} ^M \bigl( f_n |\Delta s_n|^2 + g_n |s_n|^2 \bigr),
\end{equation}
where $s_n$ and its complex conjugate ${\bar s}_n$ are considered independent -- thus there are two equations of motion, one for each sequence --  and the coefficient functions $f_n$ and $g_n$ are real. Then the Euler-Lagrange equations are
$$
g_n s_n - \Delta( f_{n + 1} \Delta s_{n + 1}) = 0
$$
(and conjugate), or
\begin{equation}
\label{EL-eqns}
	  f_{n+1} s_{n+1} - (f_{n+1} + f_n) s_n + f_n s_{n-1} = g_n s_n.
\end{equation}
As we will see below, this is the same form as the equation for the eigensequences in the LQC isotropic model. For general equations of this form, there are two solutions, since the initial data can be independently selected at two values of $n$.

\section{Conserved quantities}

Suppose that the Lagrangian ${\cal L} (n; s_n, \Delta s_n)$ is invariant under a transformation of the sequence $s_n$, i.e.
$$
	s_n \to s_n + \alpha \xi_n,
$$
but now including variations of the boundary values $s_M$ and $s_{-M}$; this will give us a conserved charge $Q_n$, as a special case of the discrete Noether's theorem. Since this is essentially what we have above in equation (\ref{parts-eqn}), then using the Euler-Lagrange equations,
\begin{equation}
\label{charge-1}
	\sum_{n = -M+1} ^M \Delta \biggl[ \frac{\partial {\cal L} (n+1; s_{n + 1}, \Delta s_{n + 1})}{\partial (\Delta s_{n + 1})} \xi_n \biggl] = 0.
\end{equation}
One way to interpret this is that a charge $Q_n$, given by
$$
	Q_n = \frac{\partial {\cal L} (n+1; s_{n + 1}, \Delta s_{n + 1})}{\partial (\Delta s_{n + 1})} \xi_n,
$$
so that $\sum \Delta Q_n = 0$. Using the Euler-Lagrange equations, we can show that $\Delta Q_n = 0$ for all $n$, not just the sum over the entire range. Therefore we have $Q_n = Q$ = constant for all $|n| \le M$.

We turn again now to the example action (\ref{iso-action}), which is invariant under $s_n \to s_n' = s_n \exp(i \alpha)$. Therefore, in the infinitesimal limit, $\xi_n = i s_n$. Since the sequences $s_n$ and ${\bar s}_n$ are considered independent, the current is
\begin{eqnarray}
\nonumber
	Q &=& \frac{\partial {\cal L}_{n+1}}{\partial (\Delta s_{n+1})} \xi_n + \frac{\partial {\cal L}_{n+1}}{\partial (\Delta {\bar s}_{n+1})} {\bar \xi}_n \\
\nonumber
	&=& i f_{n+1} \bigl[ s_n (\Delta {\bar s}_{n+1}) - {\bar s_n} (\Delta s_{n+1}) \bigr]	\\
	&=& i f_{n+1} (s_n {\bar s}_{n+1} - s_{n+1} {\bar s}_n).
\end{eqnarray}
Here, we make two comments. The first is that $Q$ depends only on the ``kinetic" coefficient function $f_n$, and not the ``potential" function $g_n$. Thus if two cosmological models have constraint equations where only the function $g_n$ differs between the two, they will have exactly the same form of conserved quantity $Q$. This occurs for Friedmann-Robertson-Walker models, where the value of the curvature constant $k$ changes only the functional form of $g_n$.
Finally, we note this is a discrete version of the conserved current for a complex scalar field. In particular, in the large parameter regime $|n| \gg 1$, we can use a Taylor series expansion of $s_{n + 1}$ and ${\bar s}_{n + 1}$, so that the sequence is approximated by a continuous function $s(n)$. This limit gives
\begin{equation}
	Q \approx i f_0 (n) \biggl[ s (n) \frac{d{\bar s} (n)}{dn} - {\bar s} (n) \frac{ds (n)}{dn} \biggr],
\end{equation}
where $f_0 (n)$ is the lowest order term in the series expansion of $f_{n + 1}$. In the context of a fundamentally discrete quantum cosmology, this $Q$ would be the semi-classical limit of this conserved quantity.

Before we turn to specific cases in isotropic LQC, we point out that a similar analysis is possible for difference equations with multiple parameters $n_i$, rather than the single parameter $n$ used here. However, this extension would require the difference equation to have constant lattice spacing. Such a constraint was developed for Bianchi I~\cite{Chi06}, although there is some debate whether physical considerations rule out this model in lieu of a difference equation with lattice refinement, i.e. the step sizes depend on the parameters $n_i$ themselves~\cite{lattice-refine}. In any case, the derivations presented above cannot be extended to such lattice refined models.

\section{Quantization of LQC isotropic model}

For the first use of the general methods developed above, we consider the flat isotropic LQC model of Ashtekar, Pawlowski and Singh (APS)~\cite{AshPawSin06b}. In this model, they consider a scalar field $\phi$ on a Friedman-Robertson-Walker space-time, and use this matter field as the internal clock for evolution. The sequences themselves are parametrized by a value $v \in \mathbb{R}$, proportional to the eigenvalue of the volume operator; the sign of $v$ gives the orientation of the spatial triad. Using the notation of APS, a sequence of values $\{\Psi (v) \}$ that solves the overall Hamiltonian constraint can be decomposed into an integral over eigensequences $e_\omega (v)$ with eigenvalues $\omega$, solving the equation
\begin{eqnarray}
\nonumber
	&& C^+ (v) e_\omega (v + 4) + C^0 (v) e_\omega (v) + C^- (v) e_\omega (v - 4) \\
\label{APS-cons}
	&& = \omega^2 B(v) e_\omega (v),
\end{eqnarray}
where
\begin{subequations}
\begin{eqnarray}
	C^+ (v) &=& \frac{3 \pi KG}{8} |v + 2| ||v+1| - |v+3||,	\\
	C^- (v) &=& C^+ (v - 4),	\\
	C^0 (v) &=& - C^+ (v) - C^- (v).
\end{eqnarray}
\end{subequations}
and
$$
	B(v) = \biggl( \frac{3}{2} \biggr)^3 K |v| \biggl | |v + 1|^{1/3} - |v - 1|^{1/3} \biggr |^3
$$
with the constant $K = 2\sqrt{3} / (3 \sqrt{3\sqrt{3}})$. The solutions $e_\omega (v)$ of the difference equation (\ref{APS-cons}) have support only on values $v = 4n + \epsilon$, where $\epsilon \in [0, 4)$. With this in mind, the relation (\ref{APS-cons}) can be put into the Euler-Lagrange form (\ref{EL-eqns}) if we let
$$
v = 4n + \epsilon,
$$
for $\epsilon \in [0, 4)$ and
$$
	s_n = e_\omega (v), \qquad
	f_n = C^- (v), \qquad
	g_n = \omega^2 B(v).
$$
Thus, these eigensequences can be derived from a Lagrangian of the form
$$
	{\cal L}_v = C^- (v) |\Delta e_\omega (v)|^2 + \omega^2 B(v) |e_\omega (v)|^2,
$$
which has a conserved quantity
\begin{equation}
\label{improv-charge}
	Q = i C^+ (v) [ e_\omega(v) {\bar e}_\omega (v + 4) - e_\omega(v + 4) {\bar e}_\omega(v)].
\end{equation}
In particular, note that $C^+ (0) \ne 0$, for those wave functions having support at the classical singularity $v = 0$; this means that the conserved quantity $Q$ is not required to be zero for those solutions. This allows the use of this quantity to relate the asymptotic linear combinations of the wave function at large values of the volume, which we examine next.

In the semi-classical limit of large spatial volumes $|v| \gg 1$, we expect to recover the usual Wheeler-deWitt (WdW) theory for this space-time. Specifically, taking the Taylor series of the difference equation (\ref{APS-cons}) at large $v$ gives a differential equation which matches the WdW equation at lowest order. These WdW solutions are worked out by APS, and the eigenfunctions are given by
\begin{equation}
\label{WdW-eigen}
	e_{\pm |k|} (v) = \frac{1}{\sqrt{2 \pi}} e^{\pm i |k| \ln |v|},
\end{equation}
where $\omega = |k| \sqrt{12 \pi G}$. When $|v| \gg 1$, the two independent solutions $e_\omega$ to the isotropic LQC difference equation, for a given $\omega$, in the semi-classical limit should each match a linear combination of the solutions $e_{\pm |k|}$ to the corresponding WdW equation:
\begin{eqnarray*}
\label{eqn-limit}
	e_\omega (v) &\to& A e_{|k|} (v) + B e_{-|k|} (v) \qquad v \gg 1, \\
	e_\omega (v) &\to& C e_{|k|} (v) + D e_{-|k|} (v) \qquad v \ll -1,
\end{eqnarray*}
for constant coefficients $\{ A, B, C, D \}$. In the large volume limit, the conserved current becomes
\begin{equation}
\label{current}
	Q \approx  \frac{3 i \pi KG}{16} |v| \biggl[ e_\omega (v) \frac{d {\bar e_\omega} (v)}{dv} - {\bar e_\omega} (v) \frac{d e_\omega}{dv}  \biggr].
\end{equation}
This asymptotic limit of $Q$ is precisely the same as the analogous conserved quantity that is obtained in WdW theory, using Noether's theorem for the appropriate continuous action.

From (\ref{WdW-eigen}) we have
\begin{equation}
	\frac{d e_\omega}{dv} = \frac{ikA}{|v|} e_{|k|} (v) - \frac{ikB}{|v|} e_{-|k|} (v).
\end{equation}
In the large positive volume limit -- i.e. large volume for a positively oriented triad --  (\ref{current}) gives
$$
	Q_{n \to \infty} = \frac{3 \pi k KG}{16} \bigl( |A|^2 - |B|^2 \bigr),
$$
while in the large negative volume limit (large volume but negative orientation of the triad),
$$
	Q_{n \to - \infty} = \frac{3 \pi k KG}{16} \bigl( |C|^2 - |D|^2 \bigr).
$$
Thus, the conservation of the current (\ref{current}) gives
\begin{equation}
\label{coeff-rel}
	|A|^2 - |B|^2 = |C|^2 - |D|^2
\end{equation}
as shown numerically\footnote{In APS, the authors choose to look only at those quantum solutions of the discrete LQC Hamiltonian constraint operator that are symmetric under the parity operator, to match with the choice of WdW eigenfunctions (\ref{WdW-eigen}). However, these sequences are constructed out of linear combinations of the eigensequences $e_\omega$, which are not symmetric under the exchange $v \to -v$. Thus, the conservation law (\ref{coeff-rel}) is non-trivial.} for their quantization~\cite{AshPawSin06a}. Note that the coefficient function $g_n$ does not enter at all into the form of the conserved quantity $Q$. Indeed, since going from the $k = 0$ isotropic model to either $k = \pm 1$ only alters this function $g_n$~\cite{FRW-models}, the discussion here goes through without change.

\section{Self-adjoint Hamiltonian constraint operators}

The last section dealt with a particular choice of factor-ordering for the Hamiltonian constraint of isotropic LQC, so it is interesting to see how generic these results are. The particular property we focus on here is the self-adjoint nature of the constraint operator; this is necessary in APS in order to find Dirac observables or physical states, chosen with a Hilbert space structure. Thus, in the following, we look for a conserved quantity $Q$ for a self-adjoint (gravitational) Hamiltonian constraint operator ${\hat H}_g$ acting on states $\psi$.

For the isotropic model, we have a basis of states $| \mu \rangle$ which are eigenstates of the volume operator, where $\mu \in \mathbb{R}$. There is a natural inner product for these states, namely,
$$
	\langle \mu | \nu \rangle = \delta_{\mu, \nu}.
$$
Note this is a Kronecker delta, not a Dirac delta function. Suppose we have an operator ${\hat A}$ acting on basis states $|\mu \rangle$ such that
$$
	{\hat A} |\mu \rangle = A^+ (\mu) |\mu + \delta \rangle + A^0 (\mu) |\mu \rangle + A^- (\mu) |\mu - \delta \rangle.
$$
where $\delta$ is a constant value indicating the step size for the lattice used in the difference equation; the value of $\delta$ results from the holonomy operator acting on the state $| \mu \rangle$, and is related to a physical choice, such as the minimum length of the model. Then, one can show that ${\hat A}$ is self-adjoint, i.e. $\langle {\hat A} \mu | \nu \rangle = \langle \mu | {\hat A} \nu \rangle$ only if
$$
	A^+ (\mu) = A^- (\mu + \delta).
$$
If we define a physical wavefunction as $\psi_\mu = \langle \psi | \mu \rangle$, then the requirement that the gravitational Hamiltonian constraint operator ${\hat H}_g$ is self-adjoint means that the constraint
$$
{\hat H} \psi_\mu = ({\hat H}_g + {\hat H}_{matt}) \psi_\mu = 0
$$
gives rise to a difference equation of the form
\begin{eqnarray}
\nonumber
	&& H^+ (\mu) \psi_{\mu + \delta} (\phi) + H^0 (\mu) \psi_\mu (\phi) + H^- (\mu) \psi_{\mu - \delta} (\phi) \\
\label{generic-eqn-1}
	&& = - {\hat H}_{matt} (\mu) \psi_\mu (\phi)
\end{eqnarray}
with the gravitational operator ${\hat H}_g$ coefficient functions $H^0 (\mu)$ and $H^- (\mu) = H^+ (\mu - \delta)$. Here, we included the possibility of one or more matter fields $\phi$ a matter Hamiltonian operator ${\hat H}_{matt}$.

For isotropic matter, we have generically that ${\hat H}_{matt}$ acts only on the matter fields $\phi$, giving at most a pre-factor $B(\mu)$ related to the volume of the space-time with triad eigenvalue $\mu$. If we use the viewpoint of APS, we use the matter field $\phi$ as an internal clock in the model, so ``evolution" means changing $\phi$. By finding solutions to the difference equation
\begin{eqnarray}
\nonumber
	&& H^+ (\mu) \psi_{\mu + \delta} (\phi) + H^0 (\mu) \psi_\mu (\phi) + H^- (\mu) \psi_{\mu - \delta} (\phi) \\
\label{generic-eqn-2}
	&& = \omega^2 B(\mu) \psi_\mu (\phi).
\end{eqnarray}
one can build up an arbitrary solution of the original constraint (\ref{generic-eqn-1}) out of linear combinations of eigensequence solutions to (\ref{generic-eqn-2}). To match up with the previous discussion, one can use a new parameter $n = \mu / \delta$ to get a lattice with unit spacing, then convert back into the physical parameter $\mu$. Thus, this equation is derivable from an action principle based on the Lagrangian
$$
	{\cal L}_\mu = H^- (\mu) |\Delta \psi_\mu (\phi)|^2 + F(\mu) |\psi_\mu (\phi)|^2
$$
where
$$
	F(\mu) = \omega^2 B(\mu) - H^0 (\mu) - H^+ (\mu) - H^- (\mu)
$$
and conserved quantity
$$
	Q = i H^+ (\mu) (\psi_\mu {\bar \psi}_{\mu + \delta} - \psi_{\mu + \delta} {\bar \psi_\mu})
$$
As commented above, the ``potential" function $F(\mu)$ is not relevant in finding the conserved quantity $Q$. Therefore, when the constraint operator ${\hat H}_g$ is self-adjoint, there is a conserved quantity $Q$ associated with the eigensequences of the operator, regardless of factor ordering issues, although specific orderings may lead to differing coefficient functions $H^+ (\mu)$. 

\section{Non-self-adjoint constraint equations in LQC}

 We have seen that a generic self-adjoint constraint equation is derivable from a discrete action, with a resulting conserved charge from the symmetry $s_n \to s'_n = s_n \exp(i \alpha)$. However, a conserved quantity exists even when the original Hamiltonian operator is not self-adjoint, but can be written in the form (\ref{generic-eqn-2}). For example, with the earlier quantization of the flat isotropic model~\cite{AshBojLew03}, we have that
\begin{eqnarray}
\label{old-cons}
\nonumber
	&& (V_{\mu + 5\mu_0} - V_{\mu + 3\mu_0}) \psi_{\mu + 4\mu_0} (\phi) \\
\nonumber
	&& - 2(V_{\mu + \mu_0} - V_{\mu - \mu_0}) \psi_\mu (\phi)	 \\
\nonumber
	&& + (V_{\mu - 3\mu_0} - V_{\mu - 5\mu_0}) \psi_{\mu - 4\mu_0} (\phi) \\
	&& = - \frac{8 \pi G \gamma^3 \mu_0^3 \ell_P^2}{3} {\hat H}_{matt} \psi_\mu (\phi),
\end{eqnarray}
where $\mu_0$ is a constant step size, related to the minimum eigenvalue of the area operator, $\gamma$ is the Immirzi parameter, $\ell_P$ the Planck length, and
\begin{equation}
\label{V-def}
	V_\mu = \biggl( \frac{\gamma |\mu|}{6} \biggr)^{3/2} \ell_P ^3
\end{equation}
are the eigenvalues of the volume operator. The precise form of the operator ${\hat H}_{matt}$ is not important, only that it acts on the matter fields $\phi$ and not directly on the triad eigenvalues $\mu$. If we write the $\phi$ dependence $\psi_\mu (\phi) = \psi_\mu \exp(i \omega \phi)$, this operator will have an action of the form
$$
	- \frac{8 \pi G \gamma^3 \mu_0^3 \ell_P^2}{3} {\hat H}_{matt} \psi_\mu (\phi) = \omega^2 B(\mu) \psi_\mu (\phi)
$$
where the function $B(\mu)$ relates to the dependence of the matter operator on the metric components. For the constraint equation (\ref{old-cons}), the step size $\delta = 4\mu_0$; the parameter $\mu$ of this function can be rescaled as $n = \mu/4 \mu_0$ to give unit lattice spacing. We define the new function
\begin{equation}
\label{s-def}
	s_n (\phi) = [V_{(4n + 1) \mu_0} - V_{(4n - 1)\mu_0}] \psi_{4 n \mu_0} (\phi),
\end{equation}
so that the constraint equation (\ref{old-cons}) is now
$$
	s_{n + 1} (\phi) - 2 s_n (\phi) + s_{n - 1} (\phi) = \frac{\omega^2 B(n) s_n (\phi)}{V_{(4n + 1) \mu_0} - V_{(4n - 1)\mu_0}}.
$$
which is of the same form as the discrete Euler-Lagrange equation (\ref{EL-eqns}) when one chooses $f_n = 1$. Note that (\ref{V-def}) and (\ref{s-def}) together requires that $s_0 = 0$, so there is no issue with evaluating the right-hand side of this derived difference equation at $n = 0$. This difference equation gives a conserved quantity, which when written in terms of the original parameter $\mu$ is
\begin{eqnarray}
\nonumber
	Q &=& i (V_{\mu + 5 \mu_0} - V_{\mu + 3 \mu_0})(V_{\mu + \mu_0} - V_{\mu - \mu_0}) \\
	&& \times [\psi_\mu (\phi) {\bar \psi}_{\mu + 4 \mu_0} (\phi) - \psi_{\mu  + 4 \mu_0} (\phi) {\bar \phi}_\mu (\phi) ]
\end{eqnarray}
The conservation of this function can be checked directly using the original difference equation (\ref{old-cons}).

A crucial difference between this conserved quantity and the charge (\ref{improv-charge}) obtained from the APS quantization is what happens to the prefactor of $Q$ at the classical singularity. As stated above, for the APS model (\ref{APS-cons}), at this singularity, $C^+ (0) \ne 0$, so $Q$ can take any real value; this allows the charge for wave functions passing through the classical singularity to be non-zero, and thus provide a relation between semi-classical limits of the wave function far away from the $v=0$ point. On the other hand, for the earlier quantization (\ref{old-cons}), we have the difference in volumes $V_{\mu + \mu_0} - V_{\mu - \mu_0}$ at $\mu = 0$ is
$$
	V_{\mu_0} - V_{-\mu_0} = 0
$$
so that the charge $Q = 0$ for {\it any} wave function passing through the $\mu = 0$ classical singularity. This results from the non-self-adjoint nature of the constraint and places a strong restriction on allowable wave functions for this quantization not present for the APS model.

As mentioned previously, $s_0 = 0$ always, so the value $\psi_0$ is undetermined for the same reason $Q = 0$ for all sequences $\psi_\mu$ -- the fact that the coefficient function of $\psi_\mu (\phi)$ in the constraint equation is zero when evaluated at $\mu = 0$. Thus, the equation (\ref{old-cons}) puts a constraint on the other values $\psi_\mu$, and thus restricts the initial data for the wave function. The space of solutions for this equation is two-dimensional; thus, the restriction on $\psi_n$ means there is a unique (up to scaling) solution to the LQC constraint, a situation known as {\it dynamical initial conditions} -- the evolution equation itself picks out the wave function without additional physical or theoretical input~\cite{Boj01}. This has the bonus that there is no requirement of choosing a specific boundary condition for the wave function, a choice which may lead to differing physical results. However, in the context of discrete equations, this may lead to radically different behavior on either side of the classical singularity $\mu = 0$~\cite{CarKha05} (specifically, the absence of pre-classical solutions for a particular orientation of the triad), although this issue has not been explored with self-adjoint constraints.

This leads to the question of whether imposing a condition $Q = 0$ for other models may lead to a similar result; strictly speaking, this is not the same as the original idea of dynamical initial conditions, since we have added a specific input to the model beyond the constraint equation, but it is worthy to look into this choice as an alternative to various boundary conditions. The answer is negative, which we now show for the APS quantization (\ref{APS-cons}). Here, the wave functions (\ref{WdW-eigen}) are chosen as an orthonormal basis of the WdW solution space, and used to find the initial data for the LQC eigensequences, but these functions are certainly not the only choice. Indeed, one can pick
\begin{eqnarray*}
	f_{1, k} (v) &\equiv& \frac{1}{\sqrt{2}} [e_{k} (v) + e_{-k} (v)] = \frac{1}{\sqrt{\pi}} \cos(|k| \ln |v|)	\\
	f_{2, k} (v) &\equiv& \frac{i}{\sqrt{2}} [e_{|k|} (v) - e_{-|k|} (v)] = \frac{1}{\sqrt{\pi}} \sin(|k| \ln |v|)
\end{eqnarray*}
For these functions, we have that $Q = 0$, so any linear combination of them also has zero charge; any eigensequence solving the quantum constraint (\ref{APS-cons}), with initial data given by such a combination for $|v| \gg 1$ will have the same zero charge. The semi-classical analysis of APS can carry forward from this point, writing the generic solution to the semi-classical differential equation using the eigenfunctions $f_{1, k}$ and $f_{2, k}$, i.e.
\begin{equation*}
	\Psi(v, \phi) = \int^\infty _{-\infty} dk [\psi_1 (k) f_{1, k} (v) e^{i \omega \phi} + \psi_2 (k) f_{2, k} (v) e^{-i \omega \phi}]
\end{equation*}
As with the APS analysis, this leads to superselection, and one can restrict solutions to only $\Psi_1$ or $\Psi_2$, where $\Psi_i$ is the solution written only in terms of the eigenfunctions $f_{i, k}$. Thus, imposing $Q = 0$ does not lead a reduction in the solution space.

\end{document}